\documentclass[aps,
  prb,
  showpacs,
  superscriptaddress,
  floatfix,
  twocolumn,
  secnumarabic,
  amssymb,
  amsmath]{revtex4}

\usepackage{graphicx}
\usepackage{bm}
\usepackage{color}
\usepackage[colorlinks=true,citecolor=blue,linkcolor=blue,pdfstartview=FitH,pdfauthor=Gross]{hyperref}

\begin{document}

\preprint{Deepak Venkateshvaran {\it et al.}, version: 20 February 2009}

\title{Epitaxial Zn$_x$Fe$_{3-x}$O$_4$ Thin Films: A Spintronic Material with Tunable Electrical and Magnetic Properties}

\author{Deepak Venkateshvaran}
 \affiliation{Walther-Mei{\ss}ner-Institut, Bayerische Akademie der
              Wissenschaften, 85748 Garching, Germany}
 \affiliation{Materials Science Research Centre, Indian Institute of Technology
                Madras, Chennai 600036, India}

\author{Matthias Althammer}
 \affiliation{Walther-Mei{\ss}ner-Institut, Bayerische Akademie der
              Wissenschaften, 85748 Garching, Germany}

\author{Andrea Nielsen}
 \affiliation{Walther-Mei{\ss}ner-Institut, Bayerische Akademie der
              Wissenschaften, 85748 Garching, Germany}

\author{Stephan Gepr\"{a}gs}
 \affiliation{Walther-Mei{\ss}ner-Institut, Bayerische Akademie der
              Wissenschaften, 85748 Garching, Germany}

\author{M.S. Ramachandra Rao}
 \affiliation{Materials Science Research Centre, Indian Institute of Technology
                Madras, Chennai 600036, India}
 \affiliation{Physics Department, Indian Institute of Technology
                Madras, Chennai 600036, India}

\author{Sebastian T. B.~Goennenwein}
 \affiliation{Walther-Mei{\ss}ner-Institut, Bayerische Akademie der
              Wissenschaften, 85748 Garching, Germany}

\author{Matthias Opel}
\email{Matthias.Opel@wmi.badw.de}
 \affiliation{Walther-Mei{\ss}ner-Institut, Bayerische Akademie der
              Wissenschaften, 85748 Garching, Germany}

\author{Rudolf Gross}
\email{Rudolf.Gross@wmi.badw.de}
 \affiliation{Walther-Mei{\ss}ner-Institut, Bayerische Akademie der
              Wissenschaften, 85748 Garching, Germany}
 \affiliation{Physik-Department, Technische Universit\"{a}t M\"{u}nchen, 85748 Garching, Germany}

\date{\today}%

\begin{abstract}
The ferrimagnetic spinel oxide Zn$_x$Fe$_{3-x}$O$_4$ combines high Curie
temperature and spin polarization with tunable electrical and magnetic
properties, making it a promising functional material for spintronic devices.
We have grown epitaxial Zn$_x$Fe$_{3-x}$O$_4$ thin films ($0\le x \le 0.9$) on
MgO(001) substrates with excellent structural properties both in pure Ar
atmosphere and an Ar/O$_2$ mixture by laser molecular beam epitaxy and
systematically studied their structural, magnetotransport and magnetic
properties. We find that the electrical conductivity and the saturation
magnetization can be tuned over a wide range ($10^2 \ldots
10^4\,\Omega^{-1}\textrm{m}^{-1}$ and $1.0 \ldots 3.2\,\mu_{\rm B}$/f.u. at
room temperature) by Zn substitution and/or finite oxygen partial pressure
during growth. Our extensive characterization of the films provides a clear
picture of the underlying physics of the spinel ferrimagnet
Zn$_x$Fe$_{3-x}$O$_4$ with antiparallel Fe moments on the $A$ and $B$
sublattice: (i) Zn substitution removes both Fe$_A^{3+}$ moments from the $A$
sublattice and itinerant charge carriers from the $B$ sublattice, (ii) growth in
finite oxygen partial pressure generates Fe vacancies on the $B$ sublattice
also removing itinerant charge carriers, and (iii) application of both Zn
substitution and excess oxygen results in a compensation effect as Zn
substitution partially removes the Fe vacancies. Both electrical conduction and
magnetism is determined by the density and hopping amplitude of the itinerant
charge carriers on the $B$ sublattice, providing electrical conduction and
ferromagnetic double exchange between the mixed-valent Fe$_B^{2+}$/Fe$_B^{3+}$
ions on the $B$ sublattice. A decrease (increase) of charge carrier density
results in a weakening (strengthening) of double exchange and thereby
a decrease (increase) of conductivity and the saturation magnetization. This
scenario is confirmed by the observation that the saturation magnetization
scales with the longitudinal conductivity. The combination of tailored
Zn$_x$Fe$_{3-x}$O$_4$ films with semiconductor materials such as ZnO in
multi-functional heterostructures seems to be particularly appealing.
\end{abstract}


\maketitle

\section{Introduction}

Spintronic materials and devices are in the focus of current research
activities \cite{Wolf:2001a,Zutic:2004a}. Regarding functional spintronic
materials, transition metal oxides are promising candidates, since they show a
rich variety of electrical and magnetic properties specifically interesting for
applications in spintronics. Evidently, useful spintronic devices such as
magnetoresistive elements based on the tunneling magnetoresistance
\cite{Moodera:1995a} or spin transistors \cite{Ohno:2000a} require
ferromagnetic materials with high Curie temperature $T_{\rm C}$ well above room
temperature and large spin polarization $P$ of the charge carriers at the Fermi
level $E_{\rm F}$. Moreover, advanced applications often require materials with
magnetic properties which can be deliberately tuned by external control
parameters such as an applied electric field \cite{Chiba:2003a}, elastic stress
\cite{Goennenwein:2008a,Bihler:2008a}, or light \cite{Tanaka:2002a}. In this
respect, Fe$_3$O$_4$ (magnetite) is a highly promising candidate. First, it has
both a high Curie temperature of $T_{\rm C} \simeq 860$\,K and according to
band structure calculations is expected a half-metal \cite{Zhang:1991}
corroborated by spin-resolved photoelectron
spectroscopy\cite{Dedkov:2002,Fonin:2007}. Second, recent experiments indicated
that both the electronic and magnetic properties of Fe$_3$O$_4$ thin films can
be nicely tailored in solid solution systems of Fe$_{3-x}M_x$O$_4$ ($M =
\textrm{Zn, Mn}$) \cite{Takaobushi:2006,Takaobushi:2007}. Third, it has been
demonstrated that the magnetic anisotropy of Fe$_3$O$_4$ can be tuned by
elastic stress imposed by a piezoelectric actuator\cite{Brandlmaier:2008a}.
Taken all together, due to their interesting, versatile, and tunable properties
magnetite thin films and heterostructures became the focus of recent research
activities.

\begin{figure}[tb]
    \includegraphics[width=\columnwidth]{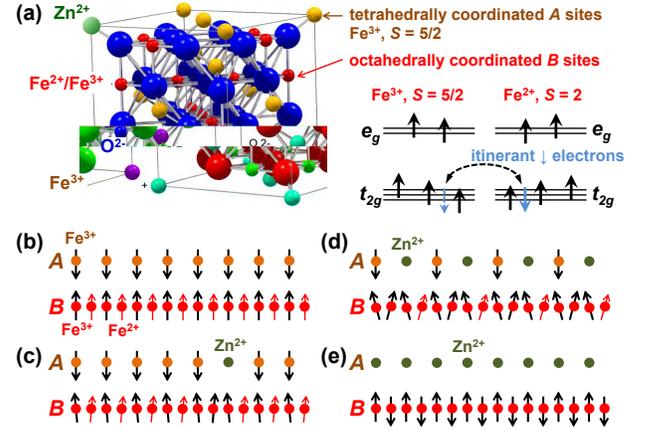}
    \caption{(color online)
             (a) Inverse spinel structure of magnetite with Fe$_A^{3+}$ on the tetrahedrally coordinated $A$ site
             and mixed-valent Fe$_B^{2+}$ ($3d^6$, S=2)/Fe$_B^{3+}$ ($3d^5, S=5/2$) ions on the octahedrally coordinated $B$ site.
             Zn substitution replaces Fe$_A^{3+}$ on the $A$ site by nonmagnetic Zn$^{2+}$ ($3d^{10}, S=0$) ions. Also
             shown is the occupation of the Fe$_B^{2+}$ and Fe$_B^{3+}$ $3d$ states with $t_{2g}$ and $e_g$ symmetry
             separated by the crystal field splitting due to the octahedral oxygen surrounding. The $t_{2g}$ spin-down
             electron can hop between Fe$_B^{2+}$ and Fe$_B^{3+}$ ions with its spin coupled anti-parallel to the local
             moment formed by the spin-up electrons owing to Hund's rule coupling.
             (b) - (e) shows the $A$ and $B$ sublattice configuration for Zn$_x$Fe$_{3-x}$O$_4$ for
             (b) $x=0$, (c) $x=1/8$, (d) $x=0.5$, and (e) $x=1$. }
    \label{fig:spinel}
\end{figure}

Before discussing the various possibilities for tuning and tailoring the
electronic and magnetic properties of magnetite we briefly review its basic
structural and magnetic properties. Magnetite is known to have an inverse
spinel structure as shown in Fig.~\ref{fig:spinel}(a). The $A$ sites (8 per
unit cell), which are surrounded by oxygen tetrahedra, are occupied by
trivalent Fe$_A^{3+}$ ions ($3d^5, S = 5/2$), whereas on the octahedrally
coordinated $B$ sites (16 per unit cell) there is an alternating arrangement of
Fe$_B^{2+}$ ($3d^6, S = 2$) and Fe$_B^{3+}$ ($3d^5, S = 5/2$) ions. Therefore,
the sum formula of magnetite can be expressed as
[Fe$^{3+}$]$_A$[Fe$^{3+}$Fe$^{2+}$]$_B$O$_4$. The density of itinerant charge
carriers is determined by the density of the $t_{2g}$ spin-down electron on the
$B$ site, i.e. by the density of Fe$_B^{2+}$. The magnetic exchange in
magnetite is governed by a combination of antiferromagnetic superexchange (SE)
and ferromagnetic double exchange (DE) interactions.  There are three
antiferromagnetic SE interactions $J_{AA}$ ($A$-O-$A$), $J_{BB}$ ($B$-O-$B$),
and $J_{AB}$ ($A$-O-$B$) between the Fe$^{3+}$ ions on the $A$ and $B$ sites mediated
by the oxygen (O) ions. In addition, there is a ferromagnetic DE interaction
mediated by the itinerant spin-down $t_{2g}$ electrons hopping between the
mixed-valent Fe ions on the $B$ sites (cf. Fig.~\ref{fig:spinel}). Owing to
Hund's rule coupling, the spins of these itinerant electrons are
antiferromagnetically coupled to the localized spins formed by the $3d$ spin-up
electrons. As pointed out by N\'{e}el \cite{Neel:1948a}, in the simplest model
ferrimagnetism in Fe$_3$O$_4$ with high $T_{\rm C}$ is obtained without any DE
interaction for large $J_{AB} \gg J_{BB}, J_{AA}$ forcing an antiparallel
alignment of the moments on the $A$ and $B$ sites (see
Fig.~\ref{fig:spinel}(b)). Since the antiparallel Fe$_A^{3+}$ and Fe$_B^{3+}$
moments compensate each other, a saturation magnetization of $4\mu_{\rm B}$/formula
unit (f.u.) is expected from the remaining Fe$_B^{2+}$ ($S=2$) moments. Later
on, the simple N\'{e}el model has been extended by Yafet and Kittel
\cite{Yafet:1952}. They proposed a more elaborate model in which the $B$
sublattice is subdivided into two Fe$_B^{2+}$ and Fe$_B^{3+}$ sublattices. It
was shown that on weakening $J_{AB}$ and strengthening $J_{BB}$, the $B$ site
magnetic moments are no longer rigidly parallel to the $A$ site moments. The
stronger $B$-O-$B$ SE interaction results in spin canting expressed by a finite
Yafet-Kittel angle and thus a reduction of the saturation magnetization. More
recent models show that a detailed modelling of the magnetic properties of
magnetite is only possible by taking into account the ferromagnetic DE
interaction of the $B$ sublattice competing with the antiferromagnetic SE
\cite{Loos:2002,McQueeney:2007,Rosencwaig:1969}.

There are several possibilities to tailor the magnetic properties of magnetite.
First, Fe$_A^{3+}$ on the $A$ site can be replaced by an isovalent nonmagnetic
ion. In this way antiferromagnetically coupled moments on the $A$ sites are
removed without affecting the magnetic exchange on the $B$ sublattices. Hence,
an increase of the saturation magnetization is expected at low substitution
level. However, on increasing the $A$ site substitution the dilution of the $A$
site moments reduces $J_{AB}$, whereas $J_{BB}$ stays about constant. Then the
moments on the $A$ and $B$ site do no longer need to be rigidly antiparallel,
resulting in a finite Yafet-Kittel angle (see Fig.~\ref{fig:spinel}(c) and
(d)). This angle increases with increasing substitution, resulting in a
reduction of the saturation magnetization. Due to the isovalent substitution
the charge carrier density should stay unaffected. However, the hopping between
the mixed-valent (Fe$_B^{2+}$/Fe$_B^{3+}$) ions on the $B$ sublattice is reduced
due to the spin canting, resulting in a lower electrical conductivity and
reduced DE interaction. To our knowledge a nonmagnetic isovalent $A$ site
substitution in Fe$_3$O$_4$ has not been reported so far. Most likely this can
be attributed to the fact that suitable trivalent ions usually prefer the
octahedral coordination on the $B$ site and thus cannot be substituted solely
on the tetrahedral $A$ site. Second, as done in our work, Fe$_A^{3+}$ on the
$A$ site can be replaced by a nonmagnetic divalent ion such as Zn$^{2+}$
($3d^{10}, S=0$) \cite{Wang:1990}. It is known that Zn$^{2+}$ preferably
occupies the tetrahedrally coordinated $A$ site in the inverse spinel structure
\cite{Harris:1996}. Again some of the antiferromagnetically coupled moments on
the $A$ sites are removed, leading to an initial increase of the saturation
magnetization at low Zn substitution \cite{Takaobushi:2006,Li:2007}. However,
in the same way as discussed before, a finite Yafet-Kittel angle and, in turn,
a reduction of the saturation magnetization is expected going to larger
substitution levels \cite{Li:2007}. Moreover, substitution of Fe$_A^{3+}$ by
Zn$^{2+}$ on the $A$ site also reduces the amount of Fe$_B^{2+}$ on the $B$
site due to charge neutrality. That is, the amount of itinerant charge carriers
mediating the DE on the $B$ sublattice is reduced as shown by photoemission
spectroscopy \cite{Takaobushi:2007}. Thus, with increasing Zn substitution the
electrical conductivity is reduced both by a reduction of the density of
itinerant charge carriers and their hopping amplitude due to spin canting.
Third, Fe vacancies can be introduced by preparing magnetite samples in excess
oxygen \cite{Wang:1990,Shepherd:1985,Aragon:1982,Aragon:1985,Shaw:2000}. Since
Fe vacancies are formally equivalent to the presence of excess of O$^{2-}$
ions, charge neutrality again requires a reduced (increased) amount of
Fe$_B^{2+}$ (Fe$_B^{3+}$) ions on the $B$ site. This strengthens
antiferromagnetic SE and weakens ferromagnetic DE on the $B$ sublattice. In
turn, this results in a finite Yafet-Kittel angle and a reduced saturation
magnetization. Furthermore, a reduced electrical conductivity is expected both
by a reduced carrier density and hopping amplitude.

\begin{figure*}[tb]
    \includegraphics[width=1.4\columnwidth]{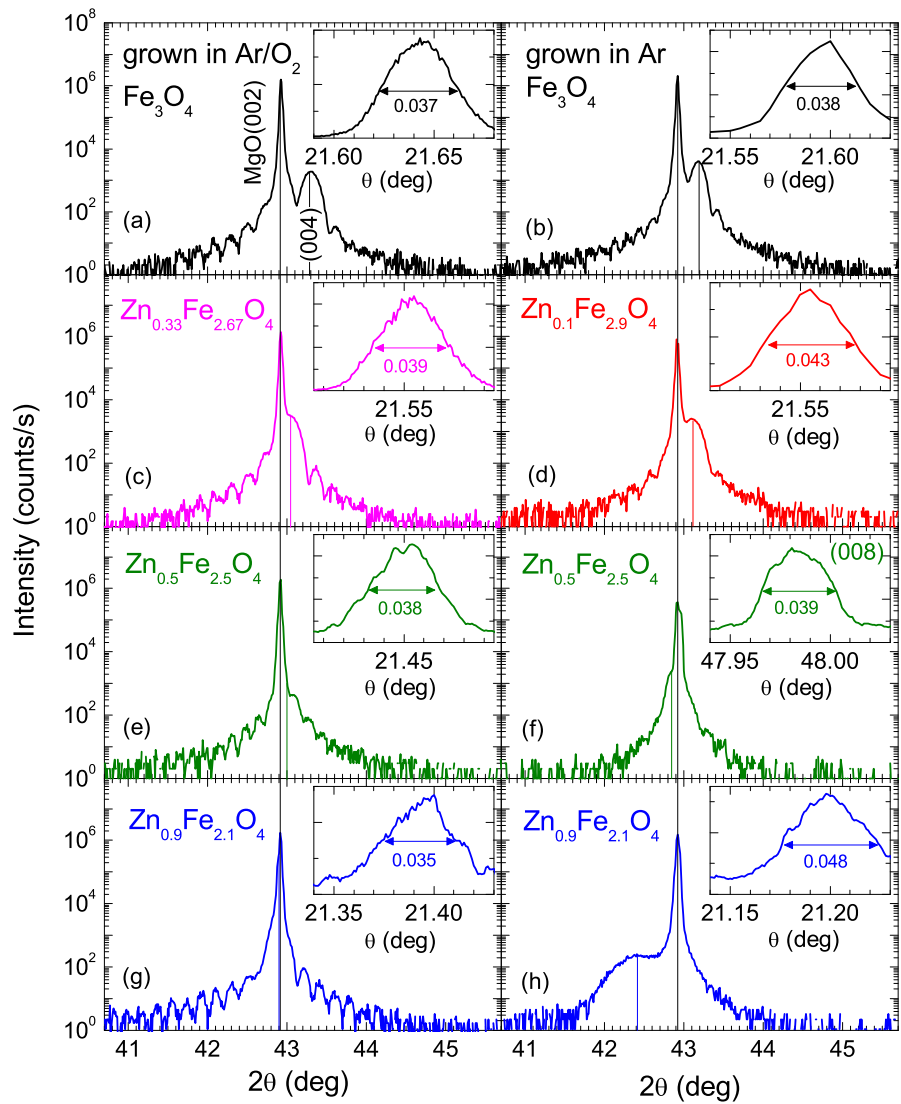}
    \caption{(color online)
             X-ray diffraction diagrams ($\omega$-$2\theta$ scans) of the out-of-plane reflections
             from Zn$_x$Fe$_{3-x}$O$_4$ thin films deposited in different growth atmospheres of Ar/O$_2$ (99:1)
             (left panels) and pure Ar (right panels). With increasing $x$ the Zn$_x$Fe$_{3-x}$O$_4$ (004) reflection shifts
             across the (002) reflection of the MgO substrates marked by the vertical
             line. The insets show the corresponding rocking curves of the Zn$_x$Fe$_{3-x}$O$_4$ (004) reflection,
             except for (f) where the Zn$_x$Fe$_{3-x}$O$_4$ (008) reflection is displayed.
             }
    \label{fig:x-ray}
\end{figure*}

Our brief discussion shows that Fe based magnetic oxides with spinel structure
are interesting materials. They do not only have high Curie temperature and
high spin polarization but also offer various opportunities to tailor their
electrical (charge carrier density, electrical conductivity) and magnetic
properties (saturation magnetization, Curie temperature). Such tunable
ferromagnetic materials are desired for spintronic devices operating at room
temperature. For example, tuning the electrical conductivity of ferromagnetic
materials with high spin polarization is promising for spin injection devices
as one can reduce and deliberately choose the conductivity mismatch between
semiconductors and ferromagnetic spin injectors. To this end, recently the
growth of epitaxial Fe$_3$O$_4$/ZnO heterostructures has been demonstrated
\cite{Nielsen:2008}. Here, we report on the growth as well as the structural,
magnetic and magnetotransport properties of epitaxial thin film samples of
Zn$_x$Fe$_{3-x}$O$_4$ with $x = 0, 0.1, 0.33, 0.5, 0.9$ deposited in different
oxygen partial pressure. We show that their saturation magnetization and
electrical conductivity can be tuned over a wide range both by Zn substitution
and varying oxygen partial pressure during growth. Furthermore, our systematic
study provides clear evidence that their electrical conductivity is closely
related to the overall magnetization.
We also carefully consider the presence of anti-phase boundaries
(APBs) in the thin film samples which could affect both their electrical
conductivity and their saturation magnetization \cite{Margulies:1996,Hibma:1999}.
However, we have clear evidence that APBs do not play a dominant role in this study.
In particular, they cannot explain the change of the magnetic and magnetotransport
properties on varying the Zn content and the deposition atmosphere. In contrast,
these properties can be consistently explained solely by 
disorder and spin canting on the $B$ sublattice, tending to
localize the itinerant charge carriers on the $B$ sublattice.

\section{Thin film growth}

Epitaxial thin films of Zn$_x$Fe$_{3-x}$O$_4$ with $x \leq 0.9$ and thicknesses
between 40\,nm and 60\,nm were deposited from stoichiometric targets on single
crystalline, (001) oriented MgO substrates by laser molecular beam
epitaxy (laser-MBE)\cite{Gross:2000}. The exact film thickness was determined by x-ray
reflectometry. Both bulk Fe$_3$O$_4$ having an inverse spinel structure
($Fd3m$) and MgO ($Fm3m$) are cubic with lattice constants of $a_{{\rm
Fe}_3{\rm O}_4}=8.394$\,{\AA} \cite{Fleet:1981} and $a_{\rm MgO}=4.212$\,{\AA},
respectively, resulting in a small lattice mismatch of $(a_{{\rm Fe}_3{\rm
O}_4} - 2a_{\rm MgO}) / 2a_{\rm MgO} = -0.4\%$. The energy density of the KrF
excimer laser ($\lambda = 248$\,nm) at the target was 3.1\,J/cm$^2$ and the
laser repetition rate 2\,Hz. At the same total pressure of $3.7 \times
10^{-3}$\,mbar, two sets of samples were deposited in two different growth
atmospheres. In pure Ar, thin films with $x=0$, 0.1, 0.5, and 0.9
were grown at a substrate temperature of $320^\circ$C. In an Ar/O$_2$ (99:1)
mixture, thin films with $x=0.5$ were deposited at $320^\circ$C and with $x=0$, 0.33,
and 0.9 at $400^\circ$C.
The growth process was monitored \textit{in-situ} by reflection high energy
electron diffraction (RHEED)\cite{Klein:1999b}. We observed four RHEED
intensity oscillations per unit cell, indicating a block-by-block growth mode
with four charge neutral blocks. This is already known for
Fe$_3$O$_4$\cite{Reisinger:2003a,Reisinger:2003b} and also holds when
substituting Zn up to Zn$_{0.9}$Fe$_{2.1}$O$_4$. A more detailed description of
the thin film deposition for Fe$_3$O$_4$ on MgO is given
elsewhere\cite{Reisinger:2003a,Reisinger:2004a}.

\section{Structural properties}

The structural properties of the Zn$_x$Fe$_{3-x}$O$_4$ films were analyzed by
high resolution x-ray diffractometry using a Bruker AXS D8 Discover four-circle
diffractometer. Our detailed analysis reveals a very high epitaxial quality of
all thin film samples. In $[00\ell]$ direction, $\omega$-$2\theta$ scans
display no impurity phases - in particular from other iron or zinc oxides. We
note that we may not be able to distinguish between Fe$_3$O$_4$ and
$\gamma$-Fe$_2$O$_3$ (maghemite) as they both share the same inverse spinel
lattice\cite{Schedin:2004}. It is known that excess oxygen results in the
formation of Fe vacancies and in the extreme case to formation of
Fe$_3$O$_4$/$\gamma$-Fe$_2$O$_3$ solid solutions
\cite{Aragon:1982,Aragon:1985}. The Fe$_3$O$_4$/$\gamma$-Fe$_2$O$_3$
thermodynamic equilibrium line was determined to $\log (p_{\rm O_2}/p_0) =
-\frac{24,634\,\textrm{K}}{T}$
with $p_0=9.1 \times 10^{13}$\,bar \cite{Aragon:1982}.
Unfortunately, both calculations and experimental studies of phase and point
defect equilibria always apply to thermal equilibrium situations which
certainly are not appropriate for the laser-MBE growth process occuring far from
equilibrium. Nevertheless, the high energy of the particles in the laser plume
can be associated to an effective growth temperature much larger than the
substrate temperature favoring the stability of Fe$_3$O$_4$
\cite{Aragon:1982,Aragon:1985}. Therefore, for the deposition temperature and
oxygen partial pressure used in our experiments we are sufficiently far away
from the Fe$_3$O$_4$/$\gamma$-Fe$_2$O$_3$ equilibrium line
\cite{Marcu:2007,Barin:1989}.

\begin{figure}[tb]
    \includegraphics[width=\columnwidth]{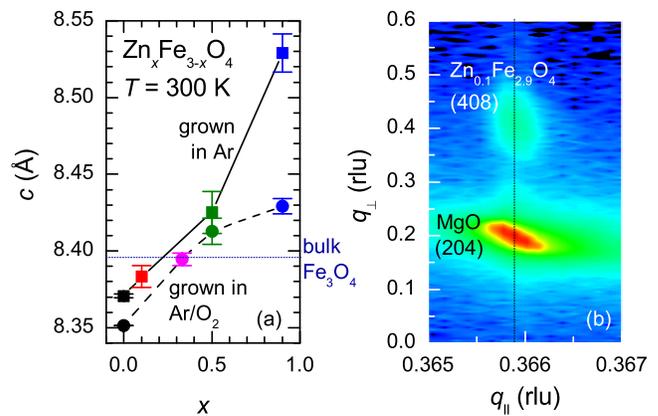}
    \caption{(color online)
             (a) Change of the $c$ axis lattice parameter of Zn$_x$Fe$_{3-x}$O$_4$ films grown on
             MgO(001) in pure Ar atmosphere (squares) and an Ar/O$_2$ mixture (circles) with Zn content $x$.
             The bulk $c$ axis lattice constant of Fe$_3$O$_4$ is marked by the dashed horizontal line.
             (b) Reciprocal space map around the (204) reflection of the MgO(001) substrate. The (408)
             reflection from the Zn$_{0.1}$Fe$_{2.9}$O$_4$ film appears at
             the same in-plane scattering vector $q_\parallel$ marked by the dashed vertical line.}
    \label{fig:rsm}
\end{figure}

As shown by Fig.~\ref{fig:x-ray}, the (004) reflection from
Zn$_x$Fe$_{3-x}$O$_4$ can be clearly observed together with satellites due to
Laue oscillations. These oscillations demonstrate that the thin films are
coherently strained and have a small surface roughness. The rocking curves of
the (004) or (008) reflections (insets of Fig.~\ref{fig:x-ray}) show a full
width at half maximum (FWHM) of $\Delta\omega \leq 0.05^\circ$. This
demonstrates the excellent structural quality of the Zn$_x$Fe$_{3-x}$O$_4$ thin
film samples with very low mosaic spread, comparable to epitaxial Fe$_3$O$_4$
films \cite{Reisinger:2003a,Reisinger:2004a}. In the $\omega$-$2\theta$ scan,
the (004) reflection from Fe$_3$O$_4$ is positioned at a slightly larger angle
than the (002) reflection from the MgO substrate (see
Fig.~\ref{fig:x-ray}(a,b)), indicating $c_{{\rm Fe}_3{\rm O}_4} < 2c_{\rm
MgO}$. With increasing Zn concentration, the (004) reflection moves to lower
angles and crosses the MgO(002) reflection at $x=0.5$ (see
Fig.~\ref{fig:x-ray}(e,f)). At $x=0.9$, the position of the (004) reflection
from the Zn$_x$Fe$_{3-x}$O$_4$ film is shifted to an angle below the (002)
reflection from MgO (see Fig.~\ref{fig:x-ray}(g,h)). This corresponds to
$c_{{\rm Zn}_{0.9}{\rm Fe}_{2.1}{\rm O}_4} > 2c_{\rm MgO}$.

The absolute value and variation of the $c$ axis lattice parameter of the
Zn$_x$Fe$_{3-x}$O$_4$ films with Zn content $x$ are related to two effects,
namely epitaxial coherency strain and the larger ionic radius of tetrahedrally
coordinated Zn$^{2+}$ of 0.6~{\AA} compared to the radius of only 0.49~{\AA} of
Fe$^{3+}$ on the tetrahedrally coordinated $A$ site \cite{Takaobushi:2006}.
First, for $x=0$ the lattice mismatch of $-0.4\%$ between Fe$_3$O$_4$ film and
MgO substrate leads to a tensile in-plane epitaxial coherency strain and, in
turn, to a slight reduction of the $c$ axis lattice parameter of the
Fe$_3$O$_4$ film below the bulk value (cf. Fig.~\ref{fig:rsm}(a)). The fully
coherent growth of the Zn$_x$Fe$_{3-x}$O$_4$ thin films is demonstrated by
reciprocal space maps around the (204) reflection of the MgO substrate (see
Fig.~\ref{fig:rsm}(b)). The (408) reflection of the film is located exactly at
the same in-plane scattering vector $q_\parallel$ as the (204) reflection of
the substrate. This clearly demonstrates that the in-plane lattice constant of
the film perfectly matches twice the one of the MgO substrate. Second, for
$x>0$ the unit cell volume of Zn$_x$Fe$_{3-x}$O$_4$ is expected to increase
about linearly with increasing $x$ due to the larger ionic radius of Zn$^{2+}$.
However, since the Zn$_x$Fe$_{3-x}$O$_4$ films grow coherently on the MgO(001)
substrate without any relaxation up to the maximal thickness of 60\,nm studied
in our experiments, the in-plane lattice constant stays unchanged. Only the
out-of-plane lattice constant is found to vary with increasing $x$. Depending
on whether the in-plane strain is tensile or compressive, a reduction or
expansion of the unit cell in $[00\ell]$ direction, respectively, is found
(Fig.~\ref{fig:rsm}(a)), resulting in a tetragonal distortion of the cubic
lattice. The $c$ axis parameter increases from a value below ($x=0$, tensile
strain) to a value above the bulk value ($x\gtrsim 0.3$, compressive strain).
The expansion of the out-of-plane lattice parameter with increasing $x$ has
been reported also for (111)-oriented films grown on Al$_2$O$_3$(0001)
substrates \cite{Takaobushi:2006}. However, we note that those films are
relaxed due to the very large lattice mismatch of 8\% between film and
substrate. Therefore, both results cannot be compared directly.

An interesting result shown in Fig.~\ref{fig:rsm} is the fact that the $c$ axis
lattice parameters of the Zn$_x$Fe$_{3-x}$O$_4$ films grown in pure Ar
atmosphere are larger than those of the films grown in an Ar/O$_2$ mixture and
also show a slightly weaker increase with Zn content in the range $0\le x \le
0.5$. This most likely is caused by the formation of Fe vacancies when growing
the films in finite oxygen partial pressure
\cite{Wang:1990,Shepherd:1985,Aragon:1982,Aragon:1985,Shaw:2000}. The presence
of Fe vacancies can be viewed as an internal negative pressure effect leading
to a reduced cell volume. The weaker increase of the $c$ axis parameter for the
films grown in Ar/O$_2$ suggests a partial compensation of the Fe vacancies by
Zn substitution. This is in agreement with the transport data discussed below.
We note, however, that this result has to be considered with some care. The
problem is that the close vicinity of film and substrate reflections in the $
\omega$-$2\theta$ scans does not allow an unambiguous derivation of the film
$c$ axis parameter with small error bars.

In summary, the structural analysis demonstrates that our
Zn$_x$Fe$_{3-x}$O$_4$(001) thin films are coherently strained and show a very
small mosaic spread. Zn substitution results in an increase of the unit cell
volume, causing the in-plane epitaxial strain to change from tensile to
compressive on increasing the Zn content $x$. Growth in excess oxygen results
in a reduction of the unit cell volume most likely due to the formation of Fe
vacancies. The latter most likely are partially compensated by additional Zn
substitution.

\section{Magnetotransport Properties}

For magnetite the electrical conduction above the Verwey transition is believed
to be determined by the hopping of the spin-down electrons between the
mixed-valent Fe$_B^{2+}$ and Fe$_B^{3+}$ ions on the $B$ sublattice. From the
hopping amplitude $t\sim 0.1$\,eV an electron conduction bandwidth $D \sim
1$\,eV is expected. However, experimental data cannot be explained within a
simple band theory. They can be better ascribed to some thermally activated
motion of charge carriers. The reason is that the bare hopping amplitude and
bandwidth is strongly reduced due to the translation of a polaronic lattice
deformation associated with electron motion\cite{Austin:1969,Mott:1970}. If the
reduced bandwidth is smaller than the energy of polarization phonons (typically
70\,meV in magnetite), small polaron hopping is expected\cite{Enim:1969}.
We note that the hopping takes place between Fe$_B^{2+}$ and Fe$_B^{3+}$ ions
on the $B$ sublattice. Below the Curie temperature, the spins of these ions are
aligned parallel in the ideal case. However, as discussed above in reality
there is some spin canting on the $B$ sublattice. This has to be taken into
account by an appropriate spin correlation factor\cite{Haubenreisser:1961}
leading to a spin dependent part of the activation energy. Since the spin
alignment is improved by applying a magnetic field, this spin-dependent part of
the activation energy can be reduced by the external field resulting in a large
negative magnetoresistance as shown below.

For magnetotransport measurements, the films were patterned into typically $45\,\mu$m
wide and $350\,\mu$m long Hall bars directed in the $\langle 010 \rangle$ direction by
photolithography and Ar ion beam milling \cite{Alff:1992a}. The longitudinal
resistivity $\rho_{xx}$ has been measured as a function of temperature $T$ and
applied magnetic field $H$ using a standard four-probe technique. The films with
$x=0.9$ were found to be insulating and are not further discussed in the following.
We note that measurements of the anomalous and the ordinary Hall effect have been
reported previously \cite{Reisinger:2004a,Venkateshvaran:2008} and are not discussed here.

\subsection{Temperature dependence of resistivity}

\begin{figure}[tb]
    \includegraphics[width=\columnwidth]{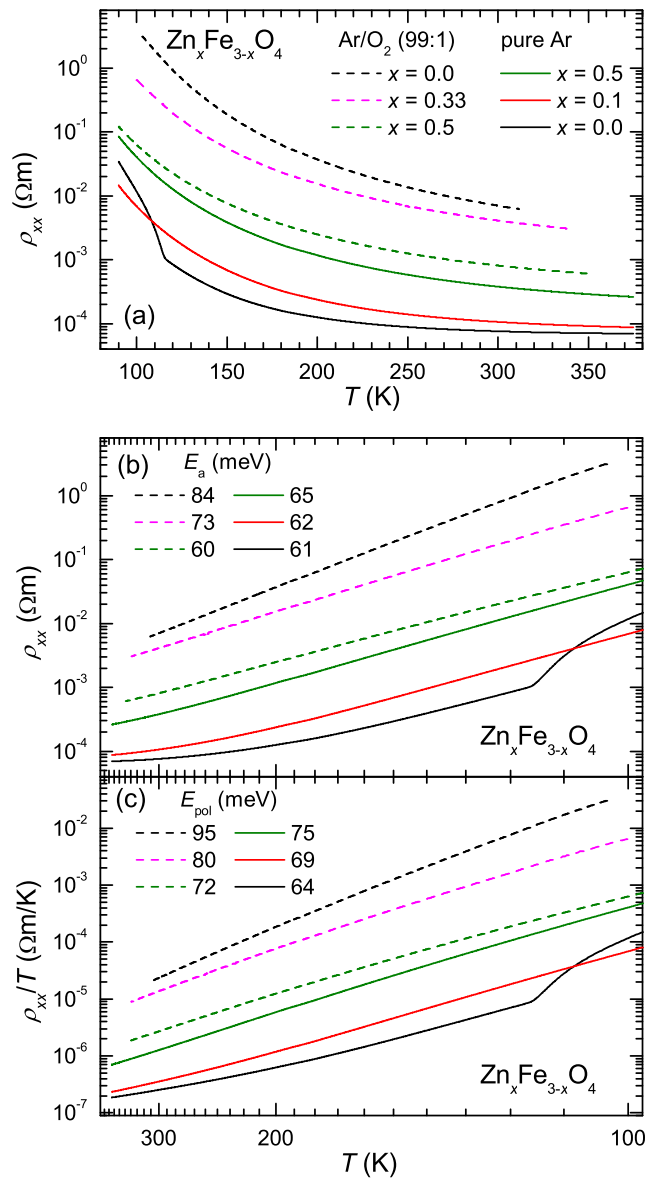}
    \caption{(color online)
             (a) Longitudinal resistivity $\rho_{xx}$ as a function of temperature $T$
             for epitaxial Zn$_x$Fe$_{3-x}$O$_4$ thin films, grown in pure Ar atmosphere
             (solid lines) or an Ar/O$_2$ (99:1) mixture (dashed lines) on MgO(001) substrates. In
             (b) and (c) $\rho_{xx}$ and $\rho_{xx}/T$ of the same samples are plotted versus a reciprocal
             temperature scale. Fitting the linear parts of the curves in (b) and (c) by
             $\rho_{xx} \propto \exp(E_{\rm a}/k_BT)$ and $\rho_{xx}/T \propto \exp(E_{\rm pol}/k_BT)$, respectively,
             gives the listed activation energies $E_{\rm a}$ and $E_{\rm pol}$. }
    \label{fig:rho}
\end{figure}

The $\rho_{xx}(T)$ curves of representative samples with $x \leq 0.5$ are shown in
Fig.~\ref{fig:rho}. Despite a similar shape of all $\rho_{xx}(T)$ curves, it is
evident that the absolute magnitude of $\rho_{xx}$ sensitively depends on the
Zn substitution $x$ and the growth atmosphere. For both samples sets grown in
pure Ar atmosphere and an Ar/O$_2$ (99:1) mixture, $\rho_{xx}(T)$ increases by
more than two orders of magnitude on decreasing $T$ from 375\,K down to 90\,K.
Plotting ${\rm log}(\rho_{xx})$ on a reciprocal $T$ scale as shown in
Fig.~\ref{fig:rho}(b) results in about linear curves over a wide temperature
range except for the $x=0$ and $x=0.1$ samples grown in pure Ar atmosphere. The
kink in $\rho_{xx}(T)$ of the $x=0$ sample results from the Verwey transition
discussed below. Fig.~\ref{fig:rho}(b) suggests that the longitudinal
resistivity of the samples with $x\gtrsim 0.1$ follows a simple activated
behavior $\rho_{xx} (T) \propto \exp(E_{\rm a}/k_{\rm B}T)$ with the
Boltzmann constant $k_{\rm B} = 1.38 \times 10^{-23}$\,J/K and
activation energies
$E_{\rm a}$ ranging between 61 and 84\,meV. However, an unambiguous
determination of the detailed transport process is difficult. As indicated by
Fig.~\ref{fig:rho}(c), the $\rho_{xx}(T)$ data also can be equally well fitted
by small polaron hopping \cite{Enim:1969,Bosman:1970,Klinger:1977}, $\rho_{xx}
(T) \propto T\exp(E_{\rm pol}/k_BT)$, with the potential barrier for polaron
hopping, $E_{\rm pol}$, ranging between 64 and 95\,meV. Actually, small
polaron hopping yields a slightly better fit for the samples grown in pure Ar
atmosphere, whereas the simple activated transport better fits the transport
data of the samples grown in an Ar/O$_2$ mixture. The derived activation
energies are typical for zinc
ferrites\cite{Wang:1990,Marcu:2007,Popandian:2002,Simsa:1988,Klinger:1977}. We
also note that a similar transport behavior with similar activation energies is
reported for the mixed-valent manganites, where charge transport is determined
by the hopping between mixed-valent Mn$^{3+}$ and Mn$^{4+}$
ions\cite{Snyder:1996,Lu:2006,Lu:2005,Lu:2000}.

We next discuss the observability of the \textit{Verwey}\cite{Verwey:1939}
transition in the measured $\rho_{xx}(T)$ curves. In bulk material, this
metal-insulator transition, associated with a cubic to monoclinic structural
phase transition, occurs at $T_V=123$\,K and in the classic picture is argued
to arise from a charge ordering of the Fe$_B^{2+}$ and Fe$_B^{3+}$ on the $B$
sublattice in a process reminiscent to Wigner crystallization \cite{Mott:1967}.
Whereas at $T > T_V$ the extra electrons from Fe$_B^{2+}$ ions can hop to
neighboring Fe$_B^{3+}$ ions on the corner sharing tetrahedral network on the
$B$ sublattice, giving rise to electrical conduction, this process is frozen in
at $T < T_V$ due to Coulomb repulsion \cite{Verwey:1947}. However, more
recently it became evident that also elastic and orbital interactions play a
significant role. This leads to a renewed controversial discussion of the
nature of the Verwey transition
\cite{McQueeney:2007,Garcia:2001,Walz:2002,Wright:2002,Leonov:2004}. It is
evident from Fig.~\ref{fig:rho} that a pronounced change in the $\rho_{xx}(T)$
curves due to the Verwey transition is observed only for the Fe$_3$O$_4$ film
($x=0$) grown in pure Ar atmosphere. For this film there is no jump but a clear
kink in the $\rho_{xx}(T)$ curve at a slightly reduced temperature of about
115\,K. For all other samples no clear signature of the Verwey transition is
observable. This can be understood in a straightforward way. As discussed
above, both substitution of Zn and introducing Fe vacancies by growing the
films in finite oxygen partial pressure results in deviations from equal
numbers of Fe$_B^{2+}$ and Fe$_B^{3+}$ on the $B$ sublattice. This can be
viewed as disorder on the $B$ sublattice, tending to smear out the Verwey
transition. The fact that the observability of the Verwey transition seems to
sensitively depend on deviations from perfect stoichiometry already has been
reported in literature
\cite{Wang:1990,Shepherd:1985,Aragon:1985,Shaw:2000,Margulies:1996} and is
further supported by our results. The other way round, we can argue that the
absence of any clear signature of the Verwey transition for the Fe$_3$O$_4$
films substituted by Zn or grown in finite oxygen partial pressure provides
evidence for the presence of an unequal number of Fe$_B^{2+}$ and Fe$_B^{3+}$
ions on the $B$ sublattice.

We next discuss the $\rho_{xx}$ values of the Zn$_x$Fe$_{3-x}$O$_4$ ($0 \le x
\le 0.5$) films grown in pure Ar atmosphere. The strong increase of resistivity
with decreasing $T$ can be understood in terms of a thermally activated hopping
type transport mechanism of the itinerant electrons on the $B$ sublattice. For
$x=0$, the room temperature $\rho_{xx}$ value ($7.6 \times 10^{-5}\,\Omega$m)
corresponds well with literature data for Fe$_3$O$_4$ thin
films\cite{Reisinger:2004a} or single crystals\cite{Todo:1995}.
In this context, we note that the presence of anti-phase
boundaries (APBs) in Fe$_3$O$_4$ thin film samples may have a significant influence on the
measured longitudinal resistivity \cite{Eerenstein:1992}. 
However, this is not the case for our samples. First, our Fe$_3$O$_4$ thin film
grown in Ar atmosphere has a room-temperature resistivity value close to that
of Fe$_3$O$_4$ single crystals ($6.2 \times 10^{-5}\,\Omega$m)
considered as APB-free \cite{Todo:1995}. Second, comparing the resistivity value
to those of Fe$_3$O$_4$ thin films with different APB density, we estimate that
the volume fraction of the APB phase in our thin film is less than 20\% \cite{Eerenstein:1992}.
Third, it is known that the formation and the density of APBs
sensitively depends on the used substrates \cite{Tiwari:2008} or the film
thickness \cite{Eerenstein:1992}. However, in our study always the same
substrate and film thickness is used. Since also Zn substitution is not
expected to severely affect the APB density, we can safely assume that
all samples grown in the same atmosphere have a similar APB density.

With increasing $x$, as discussed in the introduction, Zn$^{2+}$ preferably occupies the tetrahedrally
coordinated $A$ site in the inverse spinel structure\cite{Harris:1996}. Hence,
the substitution of Fe$_A^{3+}$ by Zn$^{2+}$ reduces the amount of Fe$_B^{2+}$ as
\begin{eqnarray}
 \left[{\rm Fe}^{3+}_{1}\right]_{A} \left[{\rm Fe}^{3+}_{1}{\rm Fe}^{2+}_{1}\right]_{B} {\rm O}^{2-}_4
 \Rightarrow \hspace*{20mm}
  \nonumber \\
  \left[{\rm Zn}^{2+}_x{\rm Fe}^{3+}_{1-x}\right]_{A} \left[{\rm Fe}^{3+}_{1+x}{\rm Fe}^{2+}_{1-x}\right]_{B} {\rm O}^{2-}_4 .
    \label{eq:Zn-substitution}
\end{eqnarray}
That is, the density of itinerant electrons on the $B$ sublattice is reduced,
resulting in an increase of the resistivity with increasing $x$. This is in
perfect agreement with our observation and literature data
\cite{Takaobushi:2006,Takaobushi:2007}. We note however, that the increase of
resistivity is not only caused by a reduction of the carrier density but also
by several other effects. First, deviations from the 1:1 balance of Fe$_B^{2+}$
and Fe$_B^{3+}$ on the $B$ sublattice causes disorder, tending to localize the
itinerant charge carriers. Second, Zn substitution dilutes the $A$ sublattice,
thereby weakening the antiferromagnetic exchange $J_{AB}$ between the $A$ and
$B$ sublattices. As discussed above, this results in spin canting on the $B$
sublattice, reducing the hopping amplitude. This amplitude is maximum for a parallel
alignment of the local moments due to strong antiparallel Hund's rule coupling
of the spins of the itinerant electrons. Both effects are expected to result in an
increase of the activation energy for the hopping transport with increasing $x$
in agreement with our observations.

The $\rho_{xx}(T)$ curves of the Zn$_x$Fe$_{3-x}$O$_4$ ($0 \le x \le 0.5$)
films grown in an Ar/O$_2$ mixture show a similar overall temperature
dependence, however, with resistivity values that are much higher than those
measured for the films grown in pure Ar atmosphere. Furthermore, the
resistivity values decrease with increasing $x$ in contrast to what is observed
for the films grown in Ar. We first discuss the origin of the much higher
resistivity value ($\rho_{xx} = 7 \times 10^{-3}\,\Omega$m at 300\,K) of the
Fe$_3$O$_4$ ($x=0$) film grown in finite oxygen partial pressure compared to
that ($\rho_{xx} = 7.6 \times 10^{-5}\,\Omega$m at 300\,K) of the film grown in
pure Ar. Since growth of magnetite in excess oxygen is known to create Fe
vacancies X$^0$ \cite{Shepherd:1985,Aragon:1982,Aragon:1985,Shaw:2000},
equivalent to two (three) missing electrons per Fe$^{2+}$ (Fe$^{3+}$) vacancy,
the requirement of charge neutrality shifts the 1:1 balance between Fe$_B^{2+}$
and Fe$_B^{3+}$ towards Fe$_B^{3+}$ as
\begin{eqnarray}
    \left[{\rm Fe}^{3+}_1\right]_{A} \left[{\rm Fe}^{3+}_{1}{\rm Fe}^{2+}_{1}\right]_B {\rm O}^{2-}_4
    \Rightarrow \hspace*{20mm}
    \nonumber \\
    \left[{\rm Fe}^{3+}_1\right]_{A} \left[{\rm Fe}^{3+}_{1+2\delta}{\rm Fe}^{2+}_{1-3\delta}{\rm X}^0_\delta \right]_{B} {\rm O}^{2-}_4 .
    \label{eq:understoichiometry}
\end{eqnarray}
Evidently, Fe vacancies in the same way as Zn substitution result in a
reduction of Fe$_B^{2+}$. Hence, the density of itinerant electrons on the $B$
sublattice decreases and, in turn, the resistivity increases with increasing
$\delta$. This is exactly what is observed in our experiments and reported in
literature \cite{Wang:1990,Parames:2006}. Actually, for Zn$_x$Fe$_{3-x}$O$_4$
($0 \le x \le 0.3$) single crystals Wang \textit{et al.} report a compositional
correspondence $x \leftrightharpoons 3\delta$ regarding the measured
resistivity data \cite{Wang:1990}. Again, with the same arguments given above
for the films grown in pure Ar, a reliable evaluation of the various mechanisms
(localization, spin canting) responsible for the increase in resistivity is not
possible. We also note that according to
eq.(\ref{eq:understoichiometry}) growth in high oxygen partial pressure (not
the case in our experiments) may lead to the formation of the cubic
$\gamma$-Fe$_2$O$_3$ (maghemite) phase in the extreme case of $\delta =
1/3$\cite{Ferguson:1958}.

Zn substitution in our magnetite films grown in an Ar/O$_2$ mixture results in
a decrease of resistivity in contrast to what is observed for the films grown
in pure Ar atmosphere. On first sight, this is astonishing and seems to be in
conflict with the above discussion, since Zn$^{2+}$ again preferably occupies
the tetrahedrally coordinated $A$ lattice sites\cite{Harris:1996}, substituting
Fe$^{3+}$ and thereby reducing carrier density. That is, one would expect an
increase of resistivity with increasing $x$. However, since Zn substitution
also results in an increase of the unit cell volume due to the larger ionic
radius of Zn$^{2+}$ compared to Fe$^{3+}$, it is expected that Zn substitution
removes part of the Fe vacancies. This is intuitive since now Fe can be more
easily incorporated into the expanded lattice. Then, it is expected that
Zn$^{2+}$ does not only remove an electron by substituting Fe$_A^{3+}$ on an
$A$ site, but also adds two/three electrons by removing a vacancy on a
Fe$_B^{2+}$/Fe$_B^{3+}$ site. That is, by Zn substitution also X$^0$ is
replaced by Fe$^{2+}$ or Fe$^{3+}$. Evidently, in total this leads to an
effective increase of the carrier density with increasing $x$ in agreement with
our experimental data in the range up to $x=0.5$. The partial removal of the Fe
vacancies by Zn substitution is in agreement with the x-ray data of
Fig.~\ref{fig:rsm}(a).

In summary, the $\rho_{xx}(T)$ curves of the Zn$_x$Fe$_{3-x}$O$_4$ films grown
under different oxygen partial pressure show a similar shape, originating from
a hopping type transport mechanism, but strongly differing absolute resistivity
values. These differences can be consistently explained by the change of the
itinerant charge carrier density on the $B$ sublattice and their hopping
amplitude by either Zn substitution or the generation of Fe vacancies due to
growth at finite oxygen partial pressure. However, applying both Zn
substitution and excess oxygen results in an increase of the carrier density as
Zn substitution is removing part of the Fe vacancies generated by excess
oxygen. It would be interesting to directly check the change of the carrier density by
measurements of the Hall effect.
However, we cannot unambiguously separate the small ordinary from the large
anomalous Hall contribution and any attempt to do so would result in large
errors of the derived carrier density \cite{Venkateshvaran:2008}.

We note that the observed dependence of the resistivities on the
Zn substitution levels $x$ cannot be simply explained by the assumption of
different APB densities in the different thin film samples. It is unlikely
that with increasing $x$ the APB density increases for samples of the same
thickness when grown in Ar atmosphere while it decreases for those prepared
in the Ar/O$_2$ mixture. Moreover, the samples show a universal scaling relation
of the anomalous Hall conductivity with the longitudinal conductivity indicating
a negligibly small APB resistivity as already pointed out earlier \cite{Venkateshvaran:2008}.

\subsection{Magnetoresistance}

\begin{figure}[tb]
\centering{\includegraphics[width=0.95\columnwidth]{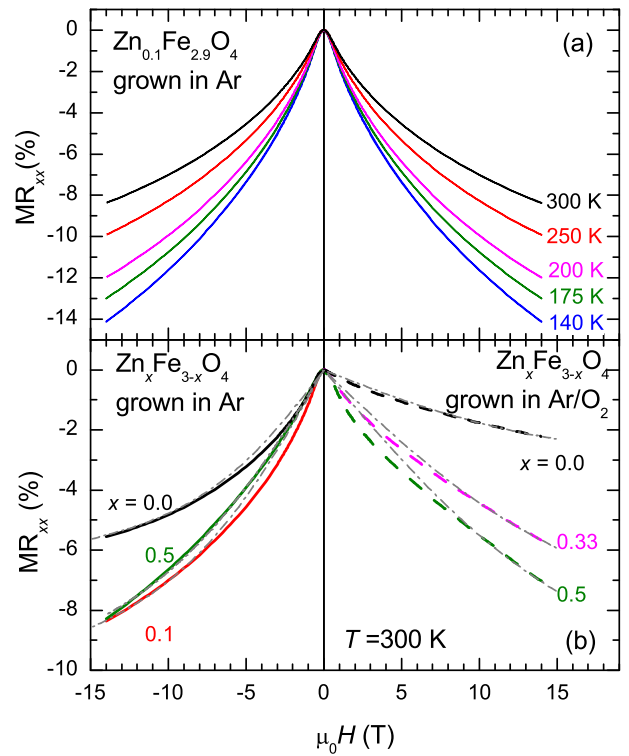}}
    \caption{(color online)
            (a) $\textrm{MR}_{xx}$ plotted versus the magnetic field $H$ applied
            perpendicular to the film plane at different temperatures for Zn$_{0.1}$Fe$_{2.9}$O$_4$
            thin films grown in pure Ar atmosphere.
            (b) $\textrm{MR}_{xx}(H)$ curves measured at 300\,K for different Zn$_x$Fe$_{3-x}$O$_4$
            grown in pure Ar atmosphere (left) or an Ar/O$_2$ mixture (right). The dash-dotted gray lines are fits to the data
            according to eq.(\ref{eq:mreffect4}).
            }
    \label{fig:MReffect}
\end{figure}

In this subsection we address the magnetoresistance (MR) of the
Zn$_x$Fe$_{3-x}$O$_4$ films. The MR effect, $\textrm{MR}_{xx} = [\rho_{xx}(H) -
\rho_{xx}(0)] / \rho_{xx}(0)$ is shown in Fig.~\ref{fig:MReffect}(a) at several
temperatures for the $x=0.1$ samples grown in pure Ar. As shown in
Fig.~\ref{fig:MReffect}(b), very similar $\textrm{MR}_{xx}(H)$ curves with
similar absolute values of $\textrm{MR}_{xx}$ at room temperature are found for
the other samples, although the absolute values of their resistivities vary by
more than two orders of magnitude (cf. Fig.~\ref{fig:rho}). This interesting
observation can be consistently explained in the framework of thermally
activated hopping of the itinerant electrons on the $B$ sublattice. Due to the
strong on-site Hund's rule coupling, the spins of the itinerant spin-down
electrons are coupled antiparallel to the localized spins formed by the spin-up
electrons. Therefore, the activation energy for the hopping process is given by
the sum of a spin independent energy ($E_{\rm a}$ or $E_{\rm pol}$) and a spin
dependent contribution $E_{\rm s}$. The latter depends on the spin canting of
the local moments on the $B$ sublattice and disappears for a perfect parallel
alignment. Since in the presence of a finite canting of the local moments an
applied magnetic field tends to improve the alignment, a magnetic field
dependent total activation energy $E(H) = E_{\rm a,pol} + E_{\rm s}(H)$ is
obtained. Then, no matter whether the transport is by thermally activated
hopping, $\rho_{xx} = \rho_{0,a} \exp[E(H)/k_BT]$, or small polaron hopping,
$\rho_{xx}/T = \rho_{0,pol} \exp[E(H)/k_BT]$, the MR effect is obtained as
\begin{eqnarray}
    \textrm{MR}_{xx} & = &  \frac{\exp \left(\frac{E_{\rm s}(H)}{k_BT} \right)-\exp \left(\frac{E_{\rm s}(0)}{k_BT} \right)}
                                            {\exp \left(\frac{E_{\rm s}(0)}{k_BT} \right)}
\nonumber \\
   & = &
   \exp \left(\frac{-\Delta E_{\rm s}(H)}{k_BT} \right) -1  \; .
    \label{eq:mreffect1}
\end{eqnarray}
with $\Delta E_{\rm s}(H) = E_{\rm s}(0)-E_{\rm s}(H)$.
Evidently, with the assumptions made, $\textrm{MR}_{xx}(H)$ is independent of
the magnitude of the prefactors ($\rho_{0,a}$ or $\rho_{0,pol}$) and the
activation energies ($E_{\rm a}$ or $E_{\rm pol}$), which may strongly vary for
samples with different $x$ and grown in different atmosphere. Moreover, similar
$\textrm{MR}_{xx}(H)$ curves with comparable absolute values are obtained, if
the magnetic field induced change $\Delta E_{\rm s}(H)$
of the spin-dependent part of the activation energy is about the same
for all samples.

Since the transport in Zn$_x$Fe$_{3-x}$O$_4$ by hopping of the $t_{2g}$
spin-down electrons between mixed-valent Fe$_B^{2+}$/Fe$_B^{3+}$ ions is
equivalent to the hopping of the $e_g$ electrons between mixed-valent
Mn$^{3+}$/Mn$^{4+}$ ions in the doped manganites, we can adopt models developed
for the MR effect in the perovskite manganites
\cite{Viret:1997a,Viret:1997b,Wagner:1998}. According to these models the
magnetic field dependent change of the activation energy for hopping between
two lattice sites $i$ and $j$ can be expressed as
\begin{eqnarray}
\Delta E_{\rm s} (H) & = &   \alpha \; \left(
    \langle \mathbf{m}_i \cdot \mathbf{m}_j \rangle_{H} -
    \langle \mathbf{m}_i \cdot \mathbf{m}_j \rangle_{H=0}
  \right)
  \; ,
    \label{eq:mreffect2}
\end{eqnarray}
where $\mathbf{m}_i$ and $\mathbf{m}_j$ are the local moments on the lattice
sites $i$ and $j$, $\langle \ldots \rangle_{H=0}$ and $\langle \ldots
\rangle_{H}$ denote the sample average at zero and finite magnetic fields,
respectively, and $\alpha$ is a constant that may depend on temperature. The
local moments $\mathbf{m}_i$ and $\mathbf{m}_j$ can be associated either with
the localized spins of the individual Fe ions or the moments of small spin
clusters. Note that these spin clusters also can be considered as spin polarons,
that is, itinerant electrons dressed by a small cloud of parallel aligned
localized spins. Evidently, $\Delta E_{\rm s}=0$ is independent of $H$, if the
local moments are already perfectly aligned in zero magnetic field (ideal
ferromagnet). In this case, no MR effect is expected. However, if there is a
finite canting of the neighboring moments, applying a magnetic field results in
a reduction of this spin canting and thereby in $\Delta E_{\rm s}>0$. According
to eq.(\ref{eq:mreffect1}) this results in a finite negative MR effect with a
magnitude determined by the value of $\Delta E_{\rm s}/k_BT$. The magnetic
field dependence of  $\Delta E_{\rm s}$ can be estimated by keeping in mind
that in the ferromagnetic state both the molecular field and the applied
magnetic field support a parallel alignment of the local moments. Since the
molecular field is usually much larger than the applied field, the deviations
of the directions of the local moments from the average direction are small. In
this case $\Delta E_{\rm s}(H)$ is found to follow the Brillouin function
$\mathfrak{B}$ giving\cite{Viret:1997a,Viret:1997b,Wagner:1998}
\begin{eqnarray}
\Delta E_{\rm s}(H) & = & \beta \; \mathfrak{B} \left( \frac{\mu_{\rm eff} \mu_0 H}{k_BT} \right)
\; .
    \label{eq:mreffect3}
\end{eqnarray}
with the vacuum permeability $\mu_0 = 4 \pi \times 10^{-7}$\,Vs/Am.
Here, $\mu_{\rm eff}$ is the average value of the local moments and $\beta$ a
constant, which may show a weak temperature dependence. Assuming $\Delta E_{\rm
s} \ll k_BT$, what certainly is a reasonable assumption at room temperature, we
can approximate eq.(\ref{eq:mreffect1}) by
\begin{eqnarray}
\textrm{MR}_{xx} & \simeq & - \beta \; \frac{\mathfrak{B} \left( \frac{\mu_{\rm eff} \mu_0 H}{k_BT} \right) }{k_BT} \; .
    \label{eq:mreffect4}
\end{eqnarray}
This shows that the magnetic field dependence of the MR effect measured at
constant temperature should follow the Brillouin function. As shown in
Fig.~\ref{fig:MReffect}(b), this is indeed the case. The data of all samples
can be well fitted by eq.(\ref{eq:mreffect4}). Fitting the data gives the
$\mu_{\rm eff}$ values of the local moments. They range between 70 to
$120\,\mu_{\rm B}$ and 60 to $70\,\mu_{\rm B}$, corresponding to about 18 to 30 and 15 to
18 Fe$^{2+}$ ions, for the samples grown in pure Ar and an Ar/O$_2$ mixture,
respectively. Here, $\mu_{\rm B}$ is Bohr's magneton. Evidently, the derived moments
are larger than the localized spins of the individual Fe ions. They can be
considered as the moments of small ferromagnetic clusters with perfectly
parallel spins, in which the electrons can move freely, but with small
misalignment between neighboring clusters. Since there are eight Fe$^{2+}$ ions
per unit cell, their size ranges between about two and four unit cells. We also
note that eq.(\ref{eq:mreffect4}) does not only describe the magnetic field
dependence of the MR effect but also nicely explains the decrease of the MR
effect with increasing temperature.

In summary, the MR effect of the Zn$_x$Fe$_{3-x}$O$_4$ films in the same way as
the $\rho_{xx}(T)$ data can be well described within a model based on thermally
activated hopping of itinerant charge carriers or small polarons between the
mixed-valent Fe$_B^{2+}$/Fe$_B^{3+}$ ions on the $B$ sublattice. The activation
energy contains a magnetic field dependent part, which depends on the
misalignment of the neighboring spin moments. Within this model,
$\textrm{MR}_{xx} \propto - \mathfrak{B}(const. \; H/T)$ is expected in good
agreement with the experimental data. In particular, the similar
$\textrm{MR}_{xx}(H)$ curves for samples with strongly different $\rho_{xx}$
values and the decrease of the MR effect with increasing temperature are nicely
reproduced.

\section{Magnetic Properties}

\begin{figure}[tb]
    \includegraphics[width=\columnwidth]{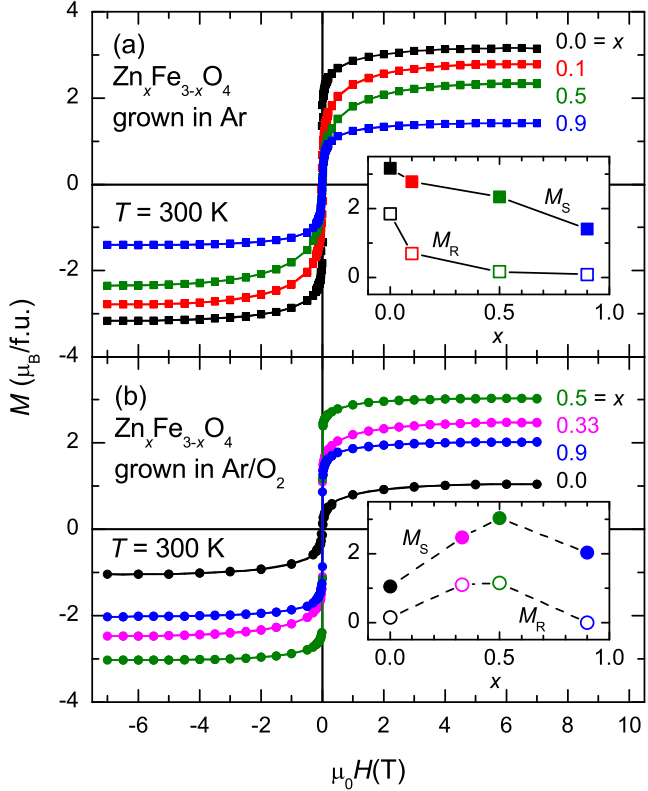}
    \caption{(color online)
             Room temperature magnetization $M$ versus magnetic field $H$ applied in the film plane
             for Zn$_x$Fe$_{3-x}$O$_4$ thin films grown in (a) pure Ar atmosphere and
             (b) in an Ar/O$_2$ (99:1) mixture. The insets show the saturation magnetization $M_{\rm S}$ (full symbols)
             and the remanence $M_{\rm R}$ (open symbols) in units of $\mu_{\rm B}$/f.u. as a function of
             the Zn substitution level $x$. The lines are guides to the eye.
             }
    \label{fig:MH}
\end{figure}

We have seen that Zn substitution and the creation of Fe vacancies in magnetite
results in changes of the carrier density and hopping amplitudes of the
itinerant charge carriers on the $B$ sublattice. Following our discussion in
the introduction, this is expected to have significant influence on the
magnetic properties. Therefore, we also systematically analyzed the magnetic
properties of the Zn$_x$Fe$_{3-x}$O$_4$ films. The magnetic characterization
was performed with the unpatterned films using SQUID magnetometry with magnetic
fields $\mu_0H$ up to 7\,T applied in the film plane. At room temperature, the
$M(H)$ loops show ferromagnetic behavior for all samples (Fig.~\ref{fig:MH}).
However, the measured saturation magnetization $M_{\rm S}$ as well as the
remanent magnetization $M_{\rm R}$ and coercive field was found to strongly
depend on both the Zn substitution level $x$ and the growth atmosphere.

We start our discussion with the stoichiometric Fe$_3$O$_4$ film grown in pure
Ar atmosphere, serving as a reference. Figure~\ref{fig:MH}(a) shows that for
this film the highest values for the room temperature saturation magnetization
($M_{\rm S} = 3.16\,\mu_{\rm B}$/f.u.) and remanence ($M_{\rm R} =
1.83\,\mu_{\rm B}$/f.u.) are obtained. We note that we do not obtain the
theoretically expected value of $M_{\rm S} = 4\,\mu_{\rm B}$/f.u. This is
typical for thin film samples due to the presence of APBs
\cite{Margulies:1996,Hibma:1999}. However, the value measured for our
film is among the highest reported in literature
and corresponds well with the volume fraction
of the APB phase of 20\% estimated above.

We next discuss the evolution of the saturation magnetization of the
Zn$_x$Fe$_{3-x}$O$_4$ films grown in pure Ar atmosphere with increasing $x$. As
shown in the inset of Fig.~\ref{fig:MH}(a), both $M_{\rm S}$ and $M_{\rm R}$
were found to decrease with increasing $x$. Recalling
eq.(\ref{eq:Zn-substitution}), the substitution of Fe$_A^{3+}$ ($3d^5, S=5/2$)
by Zn$^{2+}$ ($3d^{10}, S=0$) results in a decrease of the magnetization on the
$A$ sublattice. That is, since the compensating magnetization of the $A$
sublattice is removed, one would expect an increase of the total magnetization
of the ferrimagnet. However, Zn substitution also weakens $J_{AB}$ by diluting
the $A$ site moments. Furthermore, it converts Fe$_B^{2+}$ ($3d^6, S=2$) into
Fe$_B^{3+}$ on the $B$ sublattice, resulting in a reduction of the itinerant
charge carrier density. This weakens the ferromagnetic double exchange,
competing with antiferromagnetic superexchange interaction on the $B$
sublattice. As discussed in the introduction, taken together this leads to an
increase of the spin canting on the $B$ sublattice with increasing $x$ (cf.
Fig.~\ref{fig:spinel}), explaining the observed reduction of the total
magnetization of the ferrimagnetic Zn$_x$Fe$_{3-x}$O$_4$ films.
We note that recently Takaobushi \textit{et al.} reported a monotonic
increase of the saturation magnetization from about $0.5\,\mu_{\rm B}$/f.u. to
$3.2\,\mu_{\rm B}$/f.u. at 10\,K in the range $0 \leq x \leq 0.9$ and explained
this observation by the assumption that the $A$ sublattice magnetization
is reduced by Zn substitution \cite{Takaobushi:2006}. Comparison of the low
temperature saturation magnetization of the $x=0$ films ($3.31\,\mu_{\rm B}$/f.u.
when grown in Ar in our case, $0.5\,\mu_{\rm B}$/f.u. in Ref.~\cite{Takaobushi:2006})
clearly shows that the samples cannot be directly compared. The reason is that the films
in Ref.~\cite{Takaobushi:2006} have been grown at finite oxygen partial
pressure of $1.0 \times 10^{-6}$\,mbar. In this case, the effect of Fe vacancies
has to be taken into consideration as follows.

\begin{figure}[b]
    \includegraphics[width=\columnwidth]{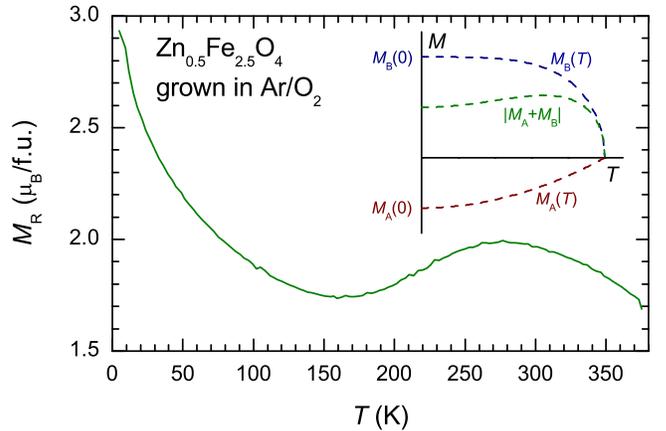}
    \caption{(color online)
             Remanent magnetization $M_{\rm R}$ plotted versus temperature $T$ for a Zn$_{0.5}$Fe$_{2.5}$O$_4$ film
             grown in an Ar/O$_2$ (99:1) mixture. The data were taken after field cooling at 7\,T on increasing the
             temperature at zero applied magnetic field. The inset schematically shows the hump in the $M(T)$
             curve as the result of two opposite sublattice magnetizations with different $T$ dependence.  }
    \label{fig:MT}
\end{figure}

The magnetic properties of the Zn$_x$Fe$_{3-x}$O$_4$ films grown in an Ar/O$_2$
(99:1) mixture are shown in Fig.~\ref{fig:MH}(b). There are pronounced
differences to the films grown in pure Ar atmosphere. First, for $x=0$ both
$M_{\rm S}$ and $M_{\rm R}$ are much lower than for the films grown in pure Ar.
Second, $M_{\rm S}$ and $M_{\rm R}$ is found to increase with increasing $x$
for $0 \leq x \leq 0.5$ and then to decrease again for $x=0.9$ as shown in the
inset of Fig.~\ref{fig:MH}(b). The coercive field decreases from about 55\,mT
to 8\,mT on increasing $x$ from 0 to 0.9. The
$M_{\rm S}(x)$ dependence can be straightforwardly explained by the presence of
Fe vacancies in films grown in finite oxygen partial pressure and their removal
by Zn substitution. For $x=0$, the Fe vacancies shift the
Fe$_B^{2+}$/Fe$_B^{3+}$ 1:1 balance towards Fe$_B^{3+}$, removing itinerant
charge carriers and weakening the ferromagnetic double exchange at the expense
of antiferromagnetic superexchange on the $B$ sublattice. This results in spin
canting on the $B$ sublattice, giving rise to a reduced saturation
magnetization. On Zn substitution, part of the Fe vacancies are removed. This
shifts the Fe$_B^{2+}$/Fe$_B^{3+}$ ratio back towards a 1:1 balance. The
related increase of the itinerant charge carrier density and DE interaction on
the $B$ sublattice reduces the spin canting with increasing $x$. This explains
the increase of the room temperature saturation magnetization in the range $0 <
x \leq 0.5$ from $M_{\rm S} = 1.05\,\mu_{\rm B}$/f.u. up to $3.02\,\mu_{\rm
B}$/f.u. at $x=0.5$. At 10\,K, $M_{\rm S}$ increases from $1.26\,\mu_{\rm
B}$/f.u. at $x=0$ up to $3.92\,\mu_{\rm B}$/f.u. at $x=0.5$. For even larger
$x$, the Fe$_A^{3+}$ moments on the $A$ sublattice are strongly diluted,
resulting in a strong weakening of the antiferromagnetic superexchange $J_{AB}$
between the $A$ and $B$ sublattices. Furthermore, Zn substitution may
overcompensate the Fe vacancies again shifting the Fe$_B^{2+}$/Fe$_B^{3+}$
balance towards Fe$_B^{3+}$. Taken together, this explains the decrease of
$M_{\rm S}$ above the critical value of $x\approx 0.5$, which of course depends
on the oxygen partial pressure during growth and the corresponding amount of Fe
vacancies. The observed increase of $M_{\rm S}$ with increasing $x$ for
Zn$_x$Fe$_{3-x}$O$_4$ films grown in an Ar/O$_2$ mixture is consistent with the
result of Takaobushi \textit{et al.}\cite{Takaobushi:2006}, who reported an
increase of $M_{\rm S}(10\,\textrm{K})$ from $0.5\,\mu_{\rm B}$/f.u. to
$3.2\,\mu_{\rm B}$/f.u. on increasing $x$ from 0 to even 0.9. However, our
systematic study strongly suggests that this increase is most likely caused by
the removal of Fe vacancies rather than due to the removal of Fe$_A^{3+}$
moments on the $A$ sites as argued in Ref.~\cite{Takaobushi:2006}.

Figure~\ref{fig:MT} shows the temperature dependence of the remanent
magnetization $M_{\rm R}$ for a Zn$_{0.5}$Fe$_{2.5}$O$_4$ film grown in an
Ar/O$_2$ mixture. In agreement with the $\rho_{xx}(T)$ data of
Fig.~\ref{fig:rho}, there is no indication for the \textit{Verwey} transition in
$M_{\rm R}(T)$ around 120\,K. This is the case for all samples, except for the Fe$_3$O$_4$
film grown in pure Ar. This observation is not only consistent with our
transport data but also with magnetization data from other groups, indicating a
smearing or suppression of the Verwey transition by an amount of Zn
substitution as small as $x = 0.01$ \cite{deTeresa:2008}.
As shown in Fig.~\ref{fig:MT}, the measured $M_{\rm R}(T)$ curve first decreases with
increasing $T$, goes through a minimum at $T \approx 150$\,K, then increases up
to $T \approx 275$\,K, before it decreases again towards room temperature.
Such a behavior is often observed in ferrimagnets and can be explained on the
basis of N\'{e}el's two sublattice model \cite{Morrish:2001}. Within this model the
origin of the hump at 275\,K is a different $T$ dependence of the two
sublattice magnetizations $M_A(T)$ and $M_B(T)$ due to different effective
molecular fields on the $A$ and $B$ sites. In our case the effective molecular
field on the $A$ sublattice is expected to be smaller due to the partial
substitution of Fe$_A^{3+}$ by nonmagnetic Zn$^{2+}$. As a result, the combined
magnetization $M(T) = M_A(T) + M_B(T)$ should show an upward hump in agreement
with the experimental result. Note that the observation of a hump, which is
directed upwards, provides direct experimental evidence for $A$ site
substitution of Zn$^{2+}$ in the inverse spinel structure \cite{Morrish:2001}.
For the $x=0.33$ sample, no hump could be observed, most likely due to a too
small imbalance in the effective molecular fields at the smaller substitution
level.

\begin{figure}[tb]
    \includegraphics[width=\columnwidth]{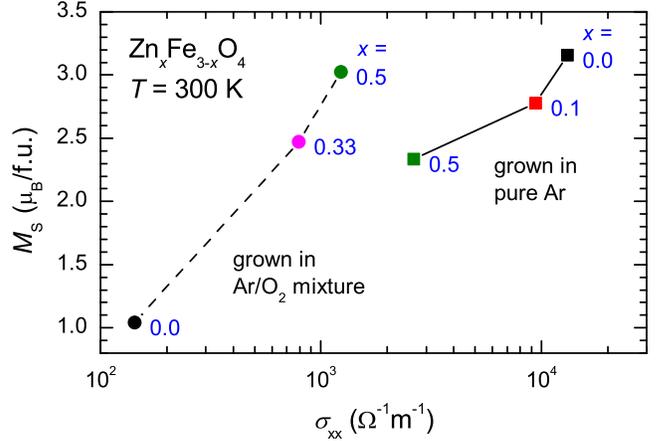}
    \caption{(color online)
             Saturation magnetization $M_{\rm S}$ plotted versus the longitudinal
             conductivity $\sigma_{xx}$ for epitaxial Zn$_x$Fe$_{3-x}$O$_4$ thin films with
             different Zn substitution $x \leq 0.5$ grown in pure Ar atmosphere (squares) or
             an Ar/O$_2$ (99:1) mixture (circles).}
    \label{fig:Spezialplot}
\end{figure}

In conclusion, the magnetic properties of the Zn$_x$Fe$_{3-x}$O$_4$ films grown
under different oxygen partial pressure strongly depend on whether the films
were grown in pure Ar atmosphere or at finite oxygen partial pressure. Fully
consistent with the transport data these differences can be explained by the
following effects: (i) For films grown in pure Ar, Zn substitution on the $A$
site removes Fe$_A^{3+}$ moments from the $A$ sublattice and due to charge
neutrality itinerant electrons from the $B$ sublattice. The resulting weakening
of both the antiferromagnetic $A$-O-$B$ SE and the ferromagnetic DE on the $B$
sublattice results in an increasing spin canting on the $B$ sublattice with
increasing $x$. This causes a reduction of the saturation magnetization with
increasing $x$. (ii) For films grown in oxygen partial pressure, the magnetic
properties are strongly influenced by the presence of Fe vacancies. They reduce
the density of itinerant electrons on the $B$ sublattice thereby weakening the
ferromagnetic DE at the expense of antiferromagnetic SE. This causes spin
canting on the $B$ sublattice and, in turn, a reduced saturation magnetization.
The effect of additional Zn substitution is the partial removal of the Fe
vacancies. This effectively increases the density of itinerant electrons on the
$B$ sublattice thereby strengthening the ferromagnetic DE. The result is an
increase of the saturation magnetization with increasing $x$.
Again, the observed dependence of the saturation magnetization
on the Zn substitution levels $x$ cannot be simply explained by different APB
densities in the different thin film samples. It is very unlikely that the APB
density increases with increasing $x$ for samples of the same thickness when grown
in Ar atmosphere while for those prepared in the Ar/O$_2$ mixture it first decreases
for $x\leq0.5$ and then increases again for $x\geq0.5$.

We finally link the magnetic properties to the transport data by plotting
the saturation magnetization $M_{\rm S}$ versus the electrical conductivity
$\sigma_{xx} \simeq 1/\rho_{xx}$\cite{Venkateshvaran:2008} for films with
various $x$ and grown in Ar or Ar/O$_2$ (Fig.~\ref{fig:Spezialplot}).
Evidently, there is a clear correlation between $M_{\rm S}$ and
$\sigma_{xx}$. Samples with high $M_{\rm S}$ have large $\sigma_{xx}$ and vice
versa. Again, this correlation can be straightforwardly explained within the
scenario presented above. Charge transport is dominated by the hopping of the
itinerant $t_{2g}$ electrons between the mixed-valent Fe$_B^{2+}$/Fe$_B^{3+}$
ions on the $B$ sublattice. Due to the strong on-site Hund's rule coupling the
spin of the hopping spin-down electron is aligned anti-parallel to the local
moment of the spin-up electrons (cf. Fig.~\ref{fig:spinel}). Therefore, the
hopping amplitude is significantly suppressed if the moments of neighboring $B$
sites are not parallel. That is, any spin canting on the $B$ sublattice
reducing the saturation magnetization also results in a reduction of the
electrical conductivity.

\section{Conclusion}

We have grown epitaxial Zn$_x$Fe$_{3-x}$O$_4$ thin films ($0\le x \le 0.9$)
with excellent structural properties both in pure Ar atmosphere and an Ar/O$_2$
mixture using laser molecular beam epitaxy. We show that the electrical
conductivity and the saturation magnetization can be tuned over a wide range
($1.0 \ldots 3.2\,\mu_{\rm B}$/f.u. and $10^2 \ldots
10^4\,\Omega^{-1}\textrm{m}^{-1}$ at room temperature) by Zn substitution
and/or finite oxygen partial pressure during growth. This demonstrates that
Zn$_x$Fe$_{3-x}$O$_4$ can be used as a material with tailored electrical and
magnetic properties. Our comprehensive study shows that electrical conduction,
magnetotransport, and the magnetic properties of Zn$_x$Fe$_{3-x}$O$_4$ are
dominated by the density and thermally activated hopping of the itinerant
$t_{2g}$ electrons between the mixed-valent Fe$_B^{2+}$/Fe$_B^{3+}$ ions,
providing electrical conduction and mediating a ferromagnetic DE interaction on
the $B$ sublattice. This scenario is confirmed by the observation that the
saturation magnetization scales with the longitudinal conductivity and the
field dependence of the MR effect. Our combined systematic analysis of the
magnetotransport and magnetic properties also provides a clear picture of the
effect of Zn substitution and growth in finite oxygen partial pressure. First,
Zn substitution removes both Fe$_A^{3+}$ moments from the $A$ sublattice and
itinerant charge carriers from the $B$ sublattice. Second, growth in finite
oxygen partial pressure generates Fe vacancies, also removing itinerant charge
carriers from the $B$ sublattice. Hence, in both cases a reduction of the charge
carrier density and a weakening of the ferromagnetic DE on the $B$ sublattice
is obtained. This results in an increase of resistivity and a reduction of the
saturation magnetization due to spin canting on the $B$ sublattice. Third,
applying both Zn substitution and growth in oxygen at the same time does not
result in an additive effect. In contrast, a compensation effect is observed,
since Zn substitution removes part of the Fe vacancies. To sum up, one can say
that the high Curie temperature together with the electrical and magnetic
properties that can be tuned over a wide range make Zn$_x$Fe$_{3-x}$O$_4$ a
promising functional material for spintronic devices. The combination of
Zn$_x$Fe$_{3-x}$O$_4$ with semiconductor materials such as ZnO in
multi-functional heterostructures \cite{Nielsen:2008} will be particularly
appealing.

\section*{Acknowledgments}

We thank Andreas Erb for the preparation of polycrystalline target materials
for the laser-MBE process and Thomas Brenninger for continuous technical
support. Financial support by the German
Science Foundation within the priority programs 1157 and 1285 (project Nos.
GR~1132/13 \& 14) and the German Excellence Initiative via the
\textit{Nanosystems Initiative Munich (NIM)} is gratefully acknowledged. D.V.
and M.S.R.R. thank the DAAD for financial support.

\end{document}